\documentclass[conference]{IEEEtran}
\IEEEoverridecommandlockouts
\usepackage{cite}
\usepackage{amsmath,amssymb,amsfonts}
\usepackage{algorithm}
\usepackage{algpseudocode}
\usepackage{graphicx,,epstopdf}
\usepackage{graphics}
\usepackage{graphicx}
\usepackage{epstopdf}
\usepackage{subcaption}
\usepackage{textcomp}
\usepackage{xcolor}
\usepackage{blindtext}
\usepackage{multirow}

\def\BibTeX{{\rm B\kern-.05em{\sc i\kern-.025em b}\kern-.08em
    T\kern-.1667em\lower.7ex\hbox{E}\kern-.125emX}}
\begin{document}

\title{Adaptive Frequency Band Selection for Accurate and Fast Positioning utilizing SOPs}
\author{Nicolas Souli, Panayiotis Kolios, and Georgios Ellinas 
\thanks{Nicolas Souli, Panayiotis Kolios, and Georgios Ellinas are with the Department of Electrical and Computer Engineering and the KIOS Research and Innovation Center of Excellence, University of Cyprus, {\tt\small \{nsouli02, pkolios, gellinas\}@ucy.ac.cy}}}

\maketitle

\begin{abstract}
Signals of opportunity (SOPs) are a promising technique that can be used for relative positioning in areas where global navigation satellite system (GNSS) information is unreliable or unavailable. This technique processes features of the various signals transmitted over a broad wireless spectrum to enable a receiver to position itself in space. This work examines the frequency selection problem in order to achieve fast and accurate positioning using only the received signal strength (RSS) of the surrounding signals. Starting with a prior belief, the problem of searching for a frequency band that best matches a predicted location trajectory is investigated. To maximize the accuracy of the position estimate, a ranking-and-selection problem is mathematically formulated. A knowledge-gradient (KG) algorithm from optimal learning theory is proposed that uses correlations in the Bayesian prior beliefs of the frequency band values to dramatically reduce the algorithm's processing time. The technique is experimentally tested for a practical scenario of an unmanned aerial vehicle (UAV) moving around a GPS-denied environment, with obtained results demonstrating its validity and practical applicability.

\end{abstract}

\begin{IEEEkeywords}
Signals of opportunity; relative positioning; vehicle tracking; adaptive learning.
\end{IEEEkeywords}

\section{Introduction}
\label{intro}
When GNSS signals are not available, various techniques have been proposed for localization, including the usage of inertial navigation systems (INS) \cite{Morales2018b}, light and range sensor measurements \cite{Maaref2018b}, RSS measurements \cite{Li2018,Ge2015}, LIDAR measurements \cite{Wang2020}, information from cellular signals \cite{Kassas2017}, signals of opportunity (SOPs)\cite{Morales,Maaref2018a,Raquet2007}, as well as a wide-range of fused information from several of the aforementioned sources. 

This work considers SOPs to calculate relative position information \cite{Morales,Maaref2018a,Raquet2007}, as SOPs are readily available and with adequate signal strength (e.g., radio, television, and cellular signals) \cite{Liu2016,Raquet2013,Morales2018b,Kapoor2017} that can be used by vehicles to navigate, especially in GNSS-challenged environments. The focus of this work is on relative positioning with the use of SOPs and inertial measurement unit (IMU) information, in areas with unreliable or not available GNSS data. 

In general, there is an extended range of real-world positioning problems where decisions are made under uncertainty and the goal is to reduce this uncertainty as much as possible. One way to accomplish this is via the gathering of as much information as possible concerning the system under investigation. One of the greatest challenges, however, is that the process of accumulating information can be very time-consuming and computationally expensive, motivating the need for conducting information gathering intelligently and efficiently~\cite{powell2012optimal,powell2010knowledge}. Thus, in this work, the aim is to explore how to achieve positioning and navigation using SOPs in an online (real-time) fashion by choosing the frequency spectrum to use for positioning (utilizing an algorithm to select the frequencies (FSA)) so as to decrease the system's computational time. The ranking-and selection problem can be utilized to model mathematically the problem at hand, i.e., to choose which frequency bands to evaluate. Further, to solve this problem, the knowledge gradient (KG) can be utilized to rank and select the best alternative frequency bands using Bayesian beliefs \cite{frazier2009knowledge}. However, the conventional implementation of a KG algorithm requires large processing and storage requirements that can render it not appropriate for a real-time implementation.

In the state of the art on different KG techniques, relatively few have focused on developing efficient online algorithms utilizing the KG approach and there are currently no works on efficient real-time KG-based algorithms for localization purposes. For other applications, such as in drug discovery, \cite{negoescu2011knowledge} utilized the KG algorithm for selecting molecular compounds. That work investigated the computational complexity of the algorithm for that particular problem and the algorithm's performance was validated using simulations. Moreover, the work in \cite{ryzhov2011bayesian} employed a Bayesian learning strategy that dealt with experiments in the energy storage sector, calculating radial basis function approximations in order to solve the parameter estimation problem.

This work examines the performance of an online KG approach for SOP-based positioning (with regards to localization accuracy and system computational time), where the beliefs about the best alternative frequency bands are computed either using linear or non-linear models. The structure of these models is utilized in order to develop computational improvements to the relative positioning system (RPS) algorithm that we previously proposed in \cite{souli2020relative} (i.e., to develop the RPS+FSA technique). In comparison to the RPS approach in \cite{souli2020relative}, the RPS+FSA algorithm manipulates the beliefs on the model parameters. Thus, since position estimation requires a large number of frequency band features, the proposed technique's (RPS+FSA) contributions are the memory and computational requirements reduction while also it enables the KG algorithm to be utilized in cases with large datasets, allowing the calculation of vehicle position estimates in real-time. 

An initial description of the frequency selection algorithm was presented in~\cite{SouliTIV2021}. This work significantly extends the description of the RPS+FSA algorithm and includes a number of experiments to ascertain how frequency selection affects the online performance of the positioning algorithm. It is demonstrated that when the number of frequency bands is large, the RPS+FSA can significantly improve the computational time, thus enabling the usage of such a technique in an real-time practical positioning scenario. In terms of baseline comparisons, the performance of RPS+FSA is compared against the measured GPS+IMU information (i.e., the ground truth) in a (real-life) practical scenario, demonstrating that the trajectory of a UAV that is computed based on SOPs in conjunction with the FSA closely matches the recorded ground truth trajectory. 

In the rest of the paper, the system model's description is included in Sec.~\ref{model}, while Sec.~\ref{KGCBnew} elaborates on the KG-based extensions on RPS for frequency selection purposes. The experimental results are presented in Sec.~\ref{results}, while Sec.~\ref{conclusion} offers concluding remarks, including future research directions.   

\section{System Model}
\label{model}
An online positioning algorithm is developed that selects the best alternative frequency bands out of $M$ alternatives and clusters them into a total of $I$ virtual transmitters that are used for the positioning of a receiver. The process is repeated over a sequence of $N$ frequency band sweeps that are required for navigating a vehicle across a specified area. The random variable $\Theta_{m}^{n}$ characterizing the $m^{th}$ alternative frequency band distribution at the $n^{th}$ spectrum sweep has an unknown mean $\mu_{m}^{n}$, known variance $\lambda$, and  covariance matrix $\Sigma_{m}^{n}$ capturing the correlation between adjacent frequency bands ($\Theta_{m}^{n} \sim N(\mu_{m}^{n},\Sigma_{m}^{n})$). The set of random variables characterizing all $M$ bands for the $n^{th}$ spectrum sweep are included in column vector $\Theta$ defined as $(\Theta_{1}^{n},...,\Theta_{M}^{n})^{T}$. 

$N$ sampling decisions, $x^{0}, x^{1}, \ldots, x^{N-1}$ are considered (one per sweep) and each measurement decision selects an alternative from set $[1,...,M]$ for testing purposes. The sample observation $\hat{y}_{m}$ is calculated and a number of iterations are required to learn the $\Theta$ values. 

Throughout the experiments, a sequence of samples are observed and their identities (decisions) are created by the sequence $x^{0},\hat{y}_{m}^{1},....,x^{n-1},\hat{y}_{m}^{n}$. Thus, the posterior distribution of $\Theta_{m}^{n+1}$ is approximated by a normal distribution with mean $\mu_{m}^{n+1}$ and covariance $\Sigma_{m}^{n+1}$.

In this work, an approximate value $p_{n}$ of the vehicle's position is obtained using a process of multilateration (Sec.~\ref{subsec:positioning}) and filtering (Sec.~\ref{subsection:EKF}), while frequency band selection using the KG policy is used to speed-up the process (Sec.~\ref{subsection:FSA}). 

\subsection{Positioning  Model}
\label{subsec:positioning}
\begin{figure}[h]
\begin{center}
\includegraphics[width=9.5cm]{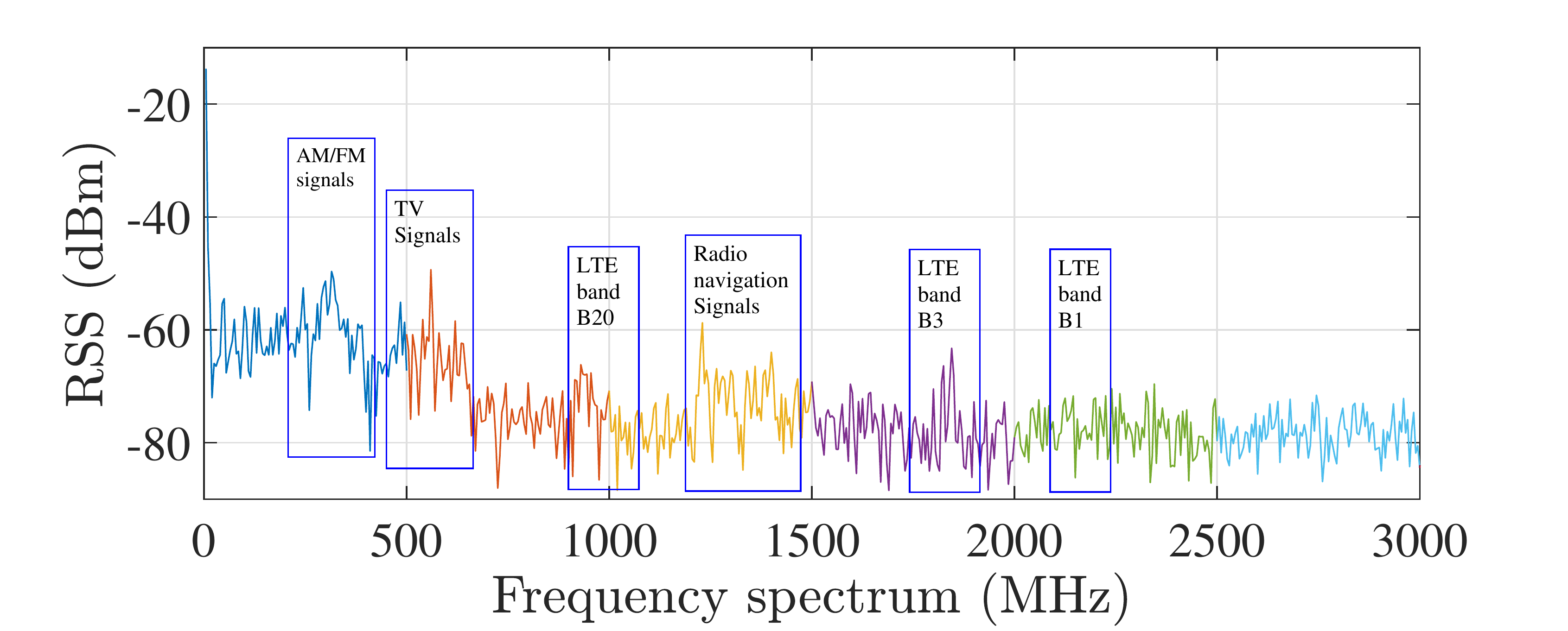}
\caption{RSS values (in dBm) from various transmission bands.}
    \label{fig:spectrumsweep}
\end{center}
\end{figure}

For the vehicle positioning problem considered in this work, a set of unknown transmitters $T_{i}$ are assumed that are allocated a subset of $M$ frequency bands, as well as a single receiver $R$ located onboard the vehicle that sweeps the wireless frequency spectrum to collect received signal strength (RSS) measurements from various transmission bands (Fig. ~\ref{fig:spectrumsweep}). The RSS measurements can present notable variances caused by various phenomena such as fading and multipath, potentially leading to erroneous calculations of the distance. In order to tackle this problem, a large number of spectrum sweeps are collected and a path-loss model is employed. 

Let $\boldsymbol{p}\epsilon \mathbb{R}^2$ denote the receiver's unknown location in the 2D plane, and $\boldsymbol{q}_{i}\epsilon \mathbb{R}^2$ denote the $i$-th transmitter's ($T_{i}$) unknown location in the 2D plane, with $i=1,...,I$. To compute distance measurements, the log-distance path-loss (PL) model is employed:
\begin{equation}
P_{L}=P_{T_{dBm}}\!\!\!-P_{R_{dBm}}=P_{L0} - 10n_{PL} \mathrm{log}_{10}(\frac{d}{d_{0}})+X_{G}
\end{equation}
with $P_{T_{dBm}}$ and $P_{R_{dBm}}$ denoting the power (in dBm) at the transmitter and receiver, respectively, $P_{L0}$ denoting the RSS at distance $d_{0}$ (i.e., the reference distance - $1$ km in this case), $n_{PL}$ denoting the path loss exponent, and $X_{G}$ denoting the log-normal shadowing term (modeled as a normal Gaussian variable with zero mean). Subsequently, to calculate $P_{L0}$ utilizing the free-space path-loss model: 
\begin{equation}
P_{L0}=20\mathrm{log}_{10}(d_0)+20\mathrm{log}_{10}(f_{c})+20\mathrm{log}_{10}(\frac{4\pi}{c_{vel}})
\label{PL0}
\end{equation}
with $f_{c}$ being the central frequency (in MHz) and $c_{vel}$ the speed of light. 
The aforementioned PL model is able to capture phenomena such as fading and multipath \cite{rappaport} that are affecting the RSS measurements, while $n_{PL}$ (i.e., the path-loss exponent) is governed by environmental conditions (e.g., in urban areas $n_{PL}=2.7-3.5$). 

Localization tries to determine an estimate of a node's position, applying range-based techniques such as trilateration or multilateration  \cite{ismail2019rssi}, \cite{Wang2013,Silva2016}. The transmitter-receiver distances are required for estimating the position of the moving vehicle, applying a range-based technique and utilizing the transmitters' estimated locations before the GNSS information was lost.  
The reader should note that the assumption in this work is that each transmitter emits information in a circle, with radius equal to the distance to the receiver, enabling the localization of the unknown node (i.e., at the intersection of these circles). Specifically, the transmitter-receiver distance (on a plane) is given by:
\begin{equation}
d_{i}=\sqrt{(p_{(1)}-q_{(1)_{i}})^{2}+(p_{(2)}-q_{(2)_{i}})^{2}}
\label{dn}
\end{equation}
where $p_{(1)}$ and $p_{(2)}$ are the position estimates of the UAV agent's receiver in the two dimensions and $q_{(1)_{i}}$, $q_{(2)_{i}}$ are the 2D coordinates of the $i$-th transmitter. 


It should be noted that the relative position of the receiver can be calculated without having prior information about the exact transmitter locations. Instead, the initial location of the transmitters can be set to any arbitrary value and the relative receiver position can be calculated thereafter based on the RSS values collected. To calculate the estimated transmitter-receiver distances an overdetermined system (with no unique solution) is utilized, linearized as $A\hat{p}=b$ with:
\begin{equation}
A=2\begin{bmatrix}
(q_{(1)_{1}}-q_{(1)_{2}}) & (q_{(2)_{1}}-q_{(2)_{2}})\\ 
...&... \\ 
 (q_{(1)_{1}}-q_{(1)_{i}})& (q_{(2)_{1}}-q_{(2)_{i}}) \\
 
\end{bmatrix}
\mathbf{\hat{p}}=\begin{bmatrix}
p_{(1)}\\ 
p_{(2)}\end{bmatrix}
\end{equation}

\begin{equation}\label{b}
b=\begin{bmatrix}
q_{(1)_{1}}^2-q_{(1)_{2}}^2+q_{(2)_{1}}^2-q_{(2)_{2}}^2+d_2^2-d_1^2\\ 
...\\ 
q_{(1)_{1}}^2-q_{(1)_{i}}^2+q_{(2)_{1}}^2-q_{(2)_{i}}^2+d_i^2-d_1^2\\ 
\end{bmatrix}.
\end{equation}
To solve this linearized system the least square (LSQ) method is utilized \cite{Brown2011,Chenshu2003}, providing the best (approximated) solution:
\begin{equation}
e=A\hat{p}-b
\label{e}
\end{equation}
\begin{equation}
\hat{\mathbf{{p}}}^{*}= \mathrm{argmin}_{{\hat{p}}}e
\label{argminxk}
\end{equation}
\begin{equation}
\hat{\mathbf{p}} = (A^{\top}A)^{-1}(A^{\top}b).
\label{eqn:LSQ}
\end{equation}

\subsection{Extended Kalman Filter}
\label{subsection:EKF}
To further improve on the position estimates, an extended Kalman filter (EKF) is further employed. Using the RSS values as well as the onboard IMU measurements of the vehicle's state at time $n$, the state at time $n+1$ can be estimated \cite{Brown2011,Wickert2018,Sazdovski2005,Li2018}. Prediction, which is the initial step, includes the $\hat{p}_{n+1}$, state transition matrix $\mathbf{F}$, and error covariance $\widehat{\mathbf{P}}_{n+1}$. Subsequently, the measurement update step calculates the Kalman gain $\mathbf{K}$ and the next error covariance matrix $\mathbf{\widehat{P}^{+}}$ as well as the next state ${\hat{p}_{n+1}^{+}}$: 
 \begin{equation}
\widehat{\mathbf{P}}_{n+1} = \mathbf{F}\widehat{\mathbf{P}}_{n}\mathbf{F}^{\top}+\mathbf{Q}
\label{phat}
\end{equation}
 \begin{equation}
\mathbf{K}=\widehat{\mathbf{P}}_{n+1}\mathbf{H^{\top}}(\mathbf{H}\widehat{\mathbf{P}}_{n+1}\mathbf{H^{\top}}+\mathbf{R})^{-1}
\label{K}
\end{equation}
 \begin{equation}
{\hat{p}_{n+1}^{+}} = \hat{p}_{n} + \mathbf{K}(z_{n+1}-\, \mathbf{H}\: \hat{p}_{n+1}) 
\label{xkhat}
\end{equation}
 \begin{equation}
\mathbf{\widehat{P}_{}}^{+}=(I-\mathbf{K}\mathbf{H})\widehat{\mathbf{P}}_{n+1} 
\label{widehatP}
\end{equation}
with $z_{n+1}$ denoting the relative measurements, $\mathbf{H}$ the measurement matrix, $\mathbf{R}$ the matrix consisting of the variances of the process noise vector, and $\mathbf{Q}$ the covariance matrix. 

\subsection{Frequency Selection Algorithm}
\label{subsection:FSA}
The EKF position estimate $p_{m}^{n}$ at time $n$ is then used for optimized frequency selection at sweep $n+1$. Let $X_{i}$ denote the contribution of $T_i$ and $\theta$ denote a vector of weights with random initial values trying to converge to the $p_{m}^{n}$'s value best estimate. 
Then, a linear model of the position estimate is given by:
 \begin{equation}
\hat{y_{m}}^{n}=\theta_{0}^{n}+X_{1}^{n}\theta_{1}^{n}+X_{2}^{n}\theta_{2}^{n}+....+X_{i}^{n}\theta{i}^{n}+\epsilon^{n}.
\label{mathematicalmodel}
\end{equation}
Since each transmitter consists of a subset of all $M$ frequency bands, then the model can be rewritten based on the features given by the measured signal strengths $RSS_{1}^{n},...,RSS_{m}^{n}$ as:
   \begin{equation}
\hat{y_{m}}^{n}=\theta_{0}^{n}+RSS_{1}^{n}\theta_{1}^{n}+RSS_{2}^{n}\theta_{2}^{n}+....+RSS_{m}^{n}\theta_{m}^{n}+\epsilon^{n}.
\label{sampledvalue}
\end{equation}
Hereafter, an independent normal prior distribution on $\theta$ and $RSS$ is assumed. Then, the mean of our belief at time $n=0$ of alternative $m$ is $\mu_{m}^{0}$ and the covariance $\Sigma_{m}^{0}$. The knowledge-gradient factor $v^{KG,n}$ captures the incremental value obtained from measuring a specific alternative $m$ as follows:
  \begin{equation}
v^{KG,n}=\mathrm{max}_{x}\mathbb{E}_{n}[\mathrm{max}_{m}\mu_{m}^{n+1}]-\mathrm{max}_{m}\mu_{m}^{n}.
\label{kgvalue}
\end{equation}

The posterior distribution $\mu_{m}^{n+1}$ is computed after each decision  $x^{n}$is made based on our prior distribution with parameters $\mu_{m}^{n}$ and $\Sigma_{m}^{n}$ as follows:
  \begin{equation}
\mu_{m}^{n+1}=\mu_{m}^{n}+\frac{\hat{y}_{m}^{n+1}-\mu_{x}^{n}}{\lambda_{x}+\Sigma_{xx}^{n}} \Sigma^{n}\epsilon_{x} 
\label{muvalue}
\end{equation}
  \begin{equation}
\Sigma_{m}^{n+1}=\Sigma_{m}^{n}+\frac{\Sigma_{m}^{n}\epsilon_{x}\epsilon_{x}^{\mathbf{\acute{}}}\Sigma_{m}^{n}}{\lambda_{x}+\Sigma_{xx}^{n}} \Sigma_{m}^{n}\epsilon_{x} 
\label{sigmavalue}
\end{equation}
where $\epsilon_{x}$ is a column $M$-vector with a value $1$ at index $x$, while the remaining values are zeros. Also, utilizing Eqs. \eqref{muvalue}, \eqref{sigmavalue}, the variance can be measured as $\tilde{\sigma}^{n}=\frac{\Sigma^{n} \epsilon_{x}}{\sqrt{\lambda_{x}+\Sigma_{xx}^{n}}} $.

Given a random variable $Z^{n+1}=\frac{\hat{y}_{m}^{n+1}-\mu_{x}^{n}}{\sqrt{\lambda_{x}+\Sigma_{xx}^{n}}}$ then the realization of the next position estimate $\mu_m^{n+1}$ can be computed as $\mu_{m}^{n+1}=\mu_{m}^{n}+\tilde{\sigma}Z^{n+1}$.
Thus, the decision $x^{n}$ on which frequency bands to sense at time $n$ can be computed as:
 \begin{equation}
\mathrm{argmax}_{x^{n}}\mathbb{E}_{n}[\mathrm{max}_{m}\mu_{m}^{n+1}]-\mathrm{max}_{m}\mu_{m}^{n} =\mathrm{argmax}h(\mu_{m}^{n},\tilde{\sigma}^{n}).
\label{XKGb}
\end{equation}

Here, $h(\mu_{m}^{n},\tilde{\sigma}^{n})$ is defined as $\mathbb{E}[\mathrm{max}_{m}g_{m}+w_{m}Z]$, where $g$ and $w$ are dimensional $M$ vectors with $g=\mu_{m}^{n}$ and $w=\tilde{\sigma}^{n}$.
To understand the calculation in Eq. \eqref{XKGb} assume a vector of consecutive choices $c$. Initially, the $w$ vector values are sorted so that $w_{1}<w_{2}<....<w_{M}$ and lines $g_{m}+w_{m}c_{m}$ are created, giving rise to:
 \begin{equation}
c_{m}=\frac{g_{m}-g_{m+1}}{w_{m+1}-w_{m}}.
\label{choicescalculation}
\end{equation}
A range of $c$ values is calculated at each iteration, with a specific choice dominating. Thus, if $c_{m+1}<c_{m}$, then the value is removed from the set until reaching the dominant choice.
%
%
The computation of the choice sequence $c$ subsequently leads to the computation of the KG value:
 \begin{equation}
v^{KG}=h(g,w)=\mathrm{log} \sum_{i=1}^{M-1}(w_{m+1}-w_{m})f(-|c_{m}|)
\label{KGvalueorh}
\end{equation}
where the function $f$ is given by $f(c)=\Phi(c)+\phi(c)$ where $\Phi$ denotes the normal cumulative distribution function and $\phi$ denotes the normal density.
Then, $v^{KG}$ is used to find the alternative $x$ values at each sample iteration and to compute the best alternative $x^{*}$ over a number of $N$ measurements. 

Regarding the KG's online learning computation, a recursive estimation variant can be realized as $\mu^{n+1}_{x}=\mu^{n}_{x}+(N-n)v^{KG,n}_{x}$.
In this non-stationary policy, frequency band selection decisions are made based on the knowledge state as well as the belief in measuring the various alternatives at an arbitrary instance $n$.

\section{KG-based RPS Extensions}
\label{KGCBnew}
Several variations to the online KG computations can be made to further improve accuracy and reduce the computational complexity. In the following sections, three such extensions are described.

\subsection{KG Algorithm for Correlated Beliefs (KGCB) on Attributes}
\label{KGCB}
To further reduce complexity in computing the KG-values a recursive expression is elaborated hereafter that is considerably more efficient than the case where $\mu_{m}^{n}$ and $\Sigma_{m}^{n}$ are maintained. Let the belief of $\Theta$ be expressed as $\Theta \sim N(X^{n}\theta^{n},X^{n}C_{m}^{n}(X^{n})^{\top})$
with a prior distribution with mean $\theta^{0}$ and covariance matrix $C_{m}^{0}$. The parameters of the prior distribution are denoted as $\mu_{m}^{0}=X\theta^{0}$ and $\Sigma_{m}^{0}=X^{n}C_{m}^{0}(X^{n})^{\top}$. Similarly, $\mu_{m}^{n}$ and $\Sigma_{m}^{n}$ are the mean and covariance following $n$ measurements and the posterior belief distribution results in a multivariate normal distribution. Let $\tilde{x^{n}}$ denote a column-vector with ones/zeros expressing the frequency band alternatives at an arbitrary instance $n$. The updated KG equations can then be expressed as follows:
 \begin{equation}
\theta^{n+1}=\theta^{n}+\frac{\hat{\epsilon}^{n+1}}{\gamma^{n}}C_{m}^{n}\tilde{x}^{n}
\label{thetarecursive}
\end{equation}
 \begin{equation}
\hat{\epsilon}^{n+1}=y_{m}^{n+1}-(X^{n}\theta^{n})^{\top}\tilde{x}^{n}
\label{epsilon}
\end{equation}
 \begin{equation}
\gamma^{n}=\lambda+(\tilde{x}^{n})^{\top}C_{m}^{n}\tilde{x^{n}}
\label{gamma}
\end{equation}
 \begin{equation}
C_{m}^{n+1}=C_{m}^{n}-\frac{1}{\gamma^{n}}(C_{m}^{n}\tilde{x}^{n}(\tilde{x}^{n})^{\top}C_{m}^{n}).
\label{covarianceeq}
\end{equation}
When the number of attributes is large, the recursive expressions for $\theta^{n}$ and $\Sigma_{m}^{n}$ are easier to maintain, since the dimensions of $\Sigma_{m}^{n}$ grow exponentially with the number of attributes. Utilizing the $\Sigma_{m}^{n}$ and $\mu_{m}^{n}$ obtained from the updated KG equations the $\tilde{\sigma}^{n}$ can then be computed as $\tilde{\sigma}^{n}=\Sigma^{n}_{x}/\sqrt{\lambda_{x}+\Sigma_{xx}^{n}}$ and can be used in the KG factor for alternative band $x$, i.e., $v^{KG}=h(\mu_{m}^{n},\tilde{\sigma})$, with $\Sigma_{x}^{n}=X_{x}^{n}C_{m}^{n}(X_{x}^{n})^{\top}$.

This approach significantly improves on the general KG algorithm because it calculates only a column vector $\Sigma_{x}^{n}$ in comparison with the full $\Sigma^{n}$ matrix.
Thus, for $x$ dimensions with $M_{x}$ alternatives, the computational complexity reduces from $\mathrm{O}(M^{2}\mathrm{log}M)$ to $\mathrm{O}(M_{x}^{2}\mathrm{log}M_{x})$. 

\subsection{KGCB Utilizing a Non-linear Model}
To further improve on the position accuracy, a non-linear model in the KG factor calculations can be employed to better capture the relationship of the correlated beliefs. Hereafter, instead of the linear relationship in Eq. \eqref{mathematicalmodel}, a quadratic function is employed as follows:
  \begin{equation}
v^{KG,n}=\mathrm{max}_{x}\mathbb{E}_{n}[\mathrm{max}_{m}J_{m}^{n+1}]-\mathrm{max}_{m}\mu_{m}^{n}
\label{kgvalue}
\end{equation}
  \begin{equation}
J_{m}^{n}=f(X)_{m}^{n}\theta^{n}+\epsilon^{n}.
\label{nonliearfunction}
\end{equation}

Let $f(X)^{n}$ be a vector of RSS values containing the information extracted from the relative transmitter $T_{i}$ and $J$ the resulting observation at time $n$. This posterior distribution is multivariate normal, having $J_{m}^{n}$ (i.e., $m$ alternative observations at each time $n$) and $\Sigma_{m}^{n}=f(X)_{m}^{n}C_{m}^{n}(f(X)_{m}^{n})^{\top}$. Under this assumption, there exist a $\theta^{n}$ and $C_{m}^{n}$ denoted with the following recursive expressions:
 \begin{equation}
\theta^{n+1}=\theta^{n}+\frac{J_{m}^{n+1}-f(\tilde{X})^{n}\theta^{n}}{\lambda_{x}+f(\tilde{X})^{n}C_{m}^{n}f(\tilde{X})^{n}}C_{m}^{n}f(\tilde{X})^{n}
\label{nonliearfunctiontheta}
\end{equation}
 \begin{equation}
C_{m}^{n+1}=C_{m}^{n}+\frac{1}{\lambda_{x}+f(\tilde{X})^{n}C_{m}^{n}f(\tilde{X})^{n}}C_{m}^{n}f(\tilde{X})^{n}(f(\tilde{X})^{n})^{\top}C_{m}^{n}.
\label{nonliearfunctioncovariave}
\end{equation}

It must be noted that $f(\tilde{X})^{n}$ denotes the RSS at the $x$ indices.
In this work, a multivariate Gaussian prior with vector $\theta^{0}$ and covariance matrix $C_{m}^{0}$ is maintained for column vector $\theta$ of unknown parameters. This gives rise to a distribution of belief on the value function $J_{m}^{n}$, $f(X)_{m}^{0}\theta^{0}$, and covariance $f(X)_{m}^{0}C_{m}^{0}(f(X)_{m}^{0})^{\top}$. Furthermore, the KG value is calculated using $\tilde{\sigma}^{n}$ as follows:
 \begin{equation}
\Sigma_{x}^{n}=f(\tilde{X})^{n}C_{m}^{n}(f(\tilde{X})^{n})^{T}
\label{newsigma}
\end{equation}
 \begin{equation}
\tilde{\sigma}^{n}=\Sigma_{x}^{n}/\sqrt{\lambda_{x}+\Sigma_{xx}^{n}}.
\label{newsigmatilde}
\end{equation}

Then, to calculate the KG value it suffices to compute $J_{m}^{n}$ and $\tilde{\sigma}^{n}$ employing $h(J_{m}^{n},\tilde{\sigma}^{n})$. Thus, the algorithm described above uses the KG concept (with the computational steps being to a great extent the same) with the non-linear model utilized for the computation of the mean and covariance values used in the KG value equation (i.e., Eq.~\eqref{kgvalue}). 

%

\subsection{Subset Policy (SP)}
To further reduce the processing time needed to compute the best choices that can maximize the position performance, a subset policy is also considered. A new set of alternatives of the vector $\mu_{m}^{n}$ is created utilizing Monte Carlo sampling which includes the most promising $K$ alternatives (e.g., in terms of RSS information) at time $n$. The reduced distribution ($\mathcal{R}$) of the remaining alternatives consists of $\mu_{K}^{\mathcal{R},n}$, and $C_{K}^{\mathcal{R},n}$. Then the KG algorithm runs using these new parameters. Thus, since $K$ is much smaller than the $M$ alternatives, the computational cost is reduced to $\mathrm{O}(K^{2}\mathrm{log}K)$.

\section{Performance Analysis}
\label{results}
\begin{figure}[h!]
\centering
\includegraphics[width=\linewidth]{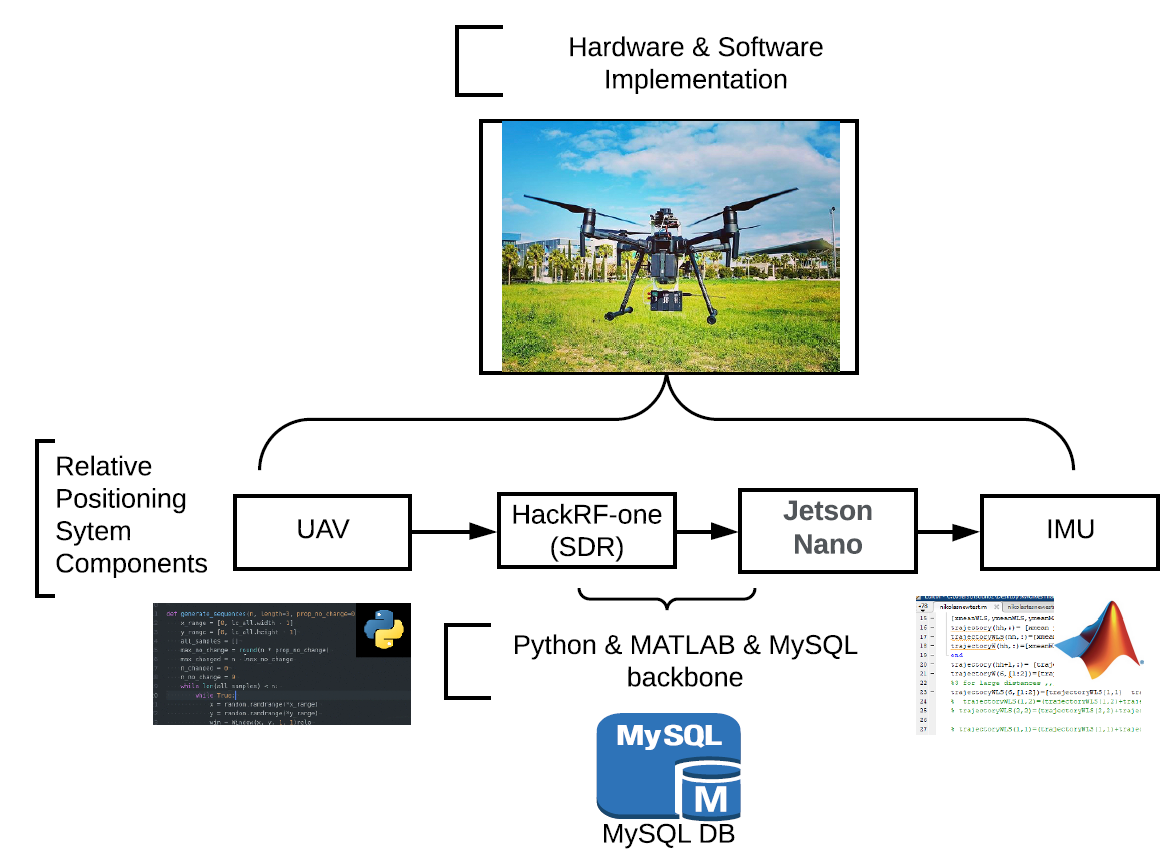}
\caption{Hardware and software implementation.}
\label{fig:UAVhardware}
\end{figure}

To evaluate the performance of the RPS+FSA system and its variants, a real setup is considered using an unmanned aerial vehicle equipped with: (1) all the onboard sensors (logging GPS and IMU measurements), (2) a software defined radio (SDR) to conduct frequency sweeps and to collect SOP RSS values ($0-3000$ MHz), (3) an Nvidia Jetson Nano to process the data, execute the RPS+FSA algorithms and provide position estimates. The overall implementation is shown in Fig. \ref{fig:UAVhardware} and was utilized for $60$ outdoor experiments conducted in a suburban area (university campus) at a constant altitude of $50$m. In the results depicted below, the performance of RPS+FSA is analyzed and compared with the UAV's GPS information fused with the IMU information that is used as the ground truth, collected over a period of $20$ minutes. A full sweep to the $M$ frequency bands occurs $U$ number of times and the RPS+FSA technique uses the best alternatives bands for positioning at an arbitrary instance in time.

\begin{figure}[t!]
\centering
\begin{subfigure}[b]{\columnwidth}
\centering
\includegraphics[width=9.5cm]{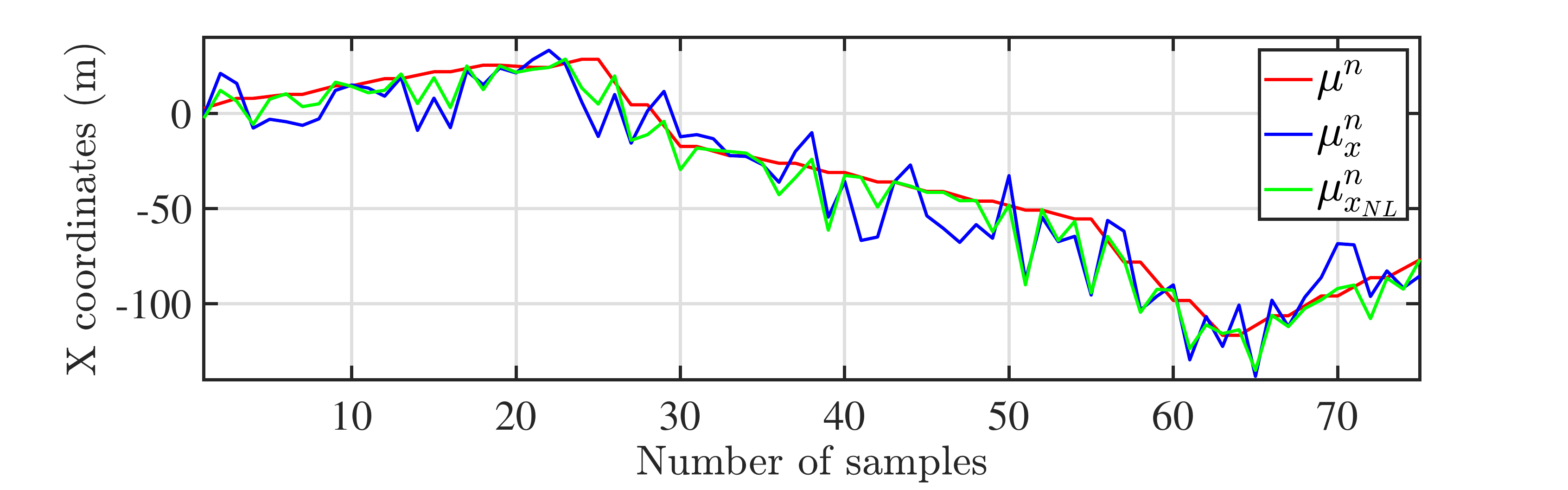}
\end{subfigure}
\begin{subfigure}[b]{\columnwidth}
\centering
\includegraphics[width=9.5cm]{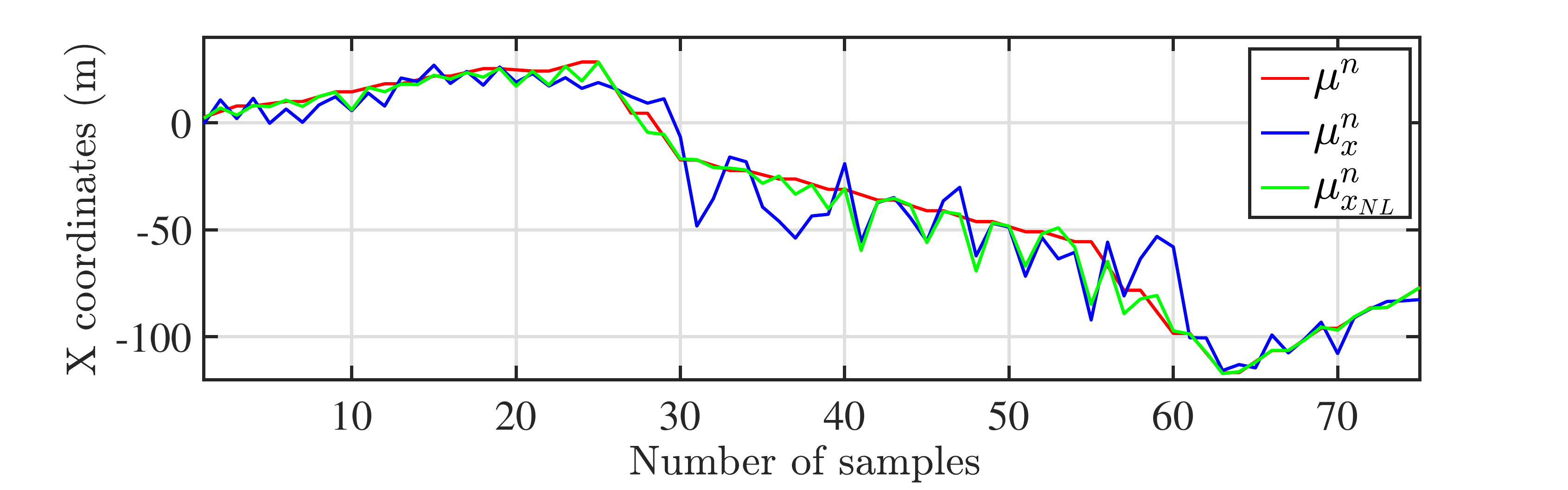}
\end{subfigure}
\begin{subfigure}[b]{\columnwidth}
\centering
\includegraphics[width=9.5cm]{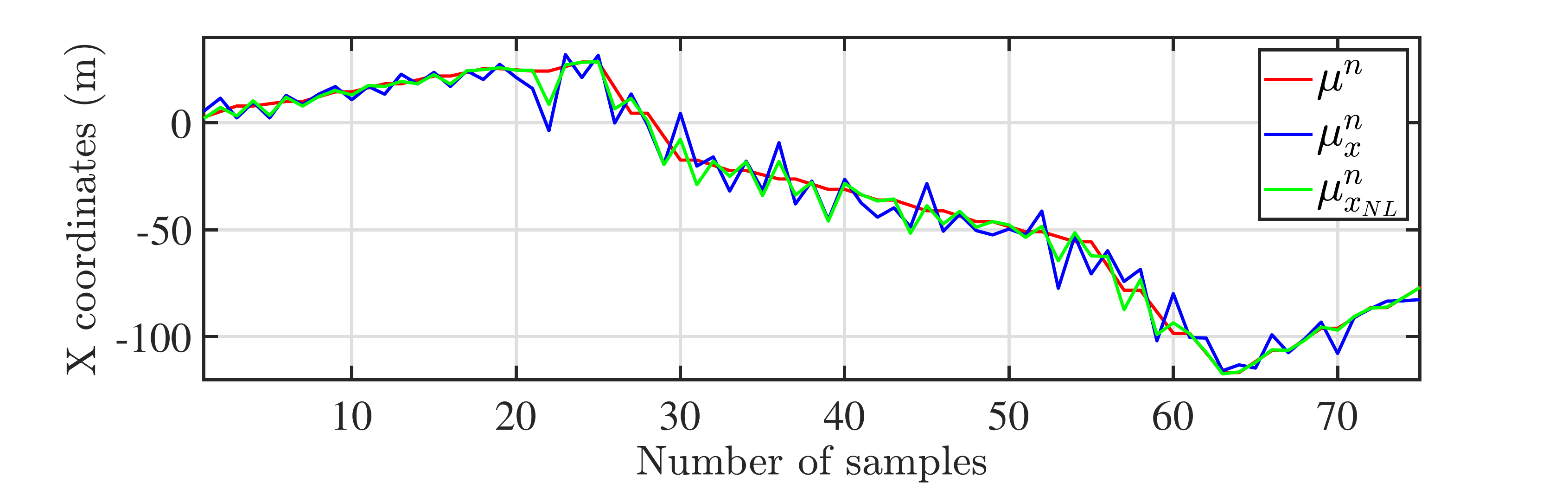}
\end{subfigure}
\caption{Estimation of $x$-coordinates for different online KGCB algorithms (top: $U=2$, middle: $U=3$, bottom: $U=4$).}
\label{fig:KGLVSNLPall}
\end{figure}

\begin{figure}[t!]
\centering
\includegraphics[width=9.5cm]{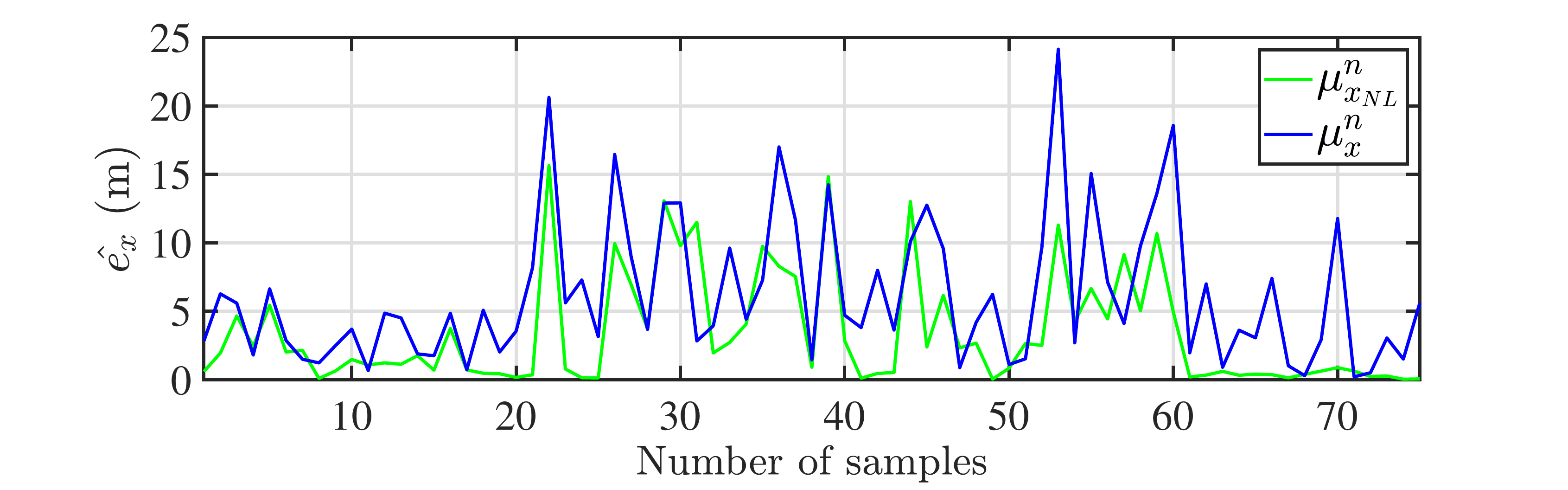}
\caption{Estimate error $\hat{e}$ for the $x$-coordinates when $U=4$.}
\label{fig:errorXYNL}
\end{figure}

Figure~\ref{fig:KGLVSNLPall} depicts the $x$-coordinate GPS+IMU positions (in red), and the position estimates obtained by the online KGCB algorithms (linear model (in blue) and non-linear model (in green)) for different values of $U$, while the estimate error $\hat{e}$ for the $x$-coordinates (for the non-linear and linear models) is shown in Fig. \ref{fig:errorXYNL}. Note that the $y$-coordinate error values follow the same pattern and thus are omitted due to space limitations.

As shown in Fig.~\ref{fig:KGLVSNLPall}, by frequently revisiting all frequency bands (i.e., higher $U$ values) the algorithm is able to better track the vehicle movement over time. In addition, both the linear and non-linear KGCB models achieve adequate performance. Further, as shown in Fig. \ref{fig:errorXYNL}, the error observed using the linear model reaches up to $20$m in some cases, while the non-linear model is able to reduce the error considerably (on average), as the non-linear parameter estimation manages to better approximate the weights and thus achieve better performance results compared to its linear counterpart.

\begin{figure}[t!]
\centering
\begin{subfigure}[b]{\columnwidth}
\centering
\includegraphics[width=9.5cm]{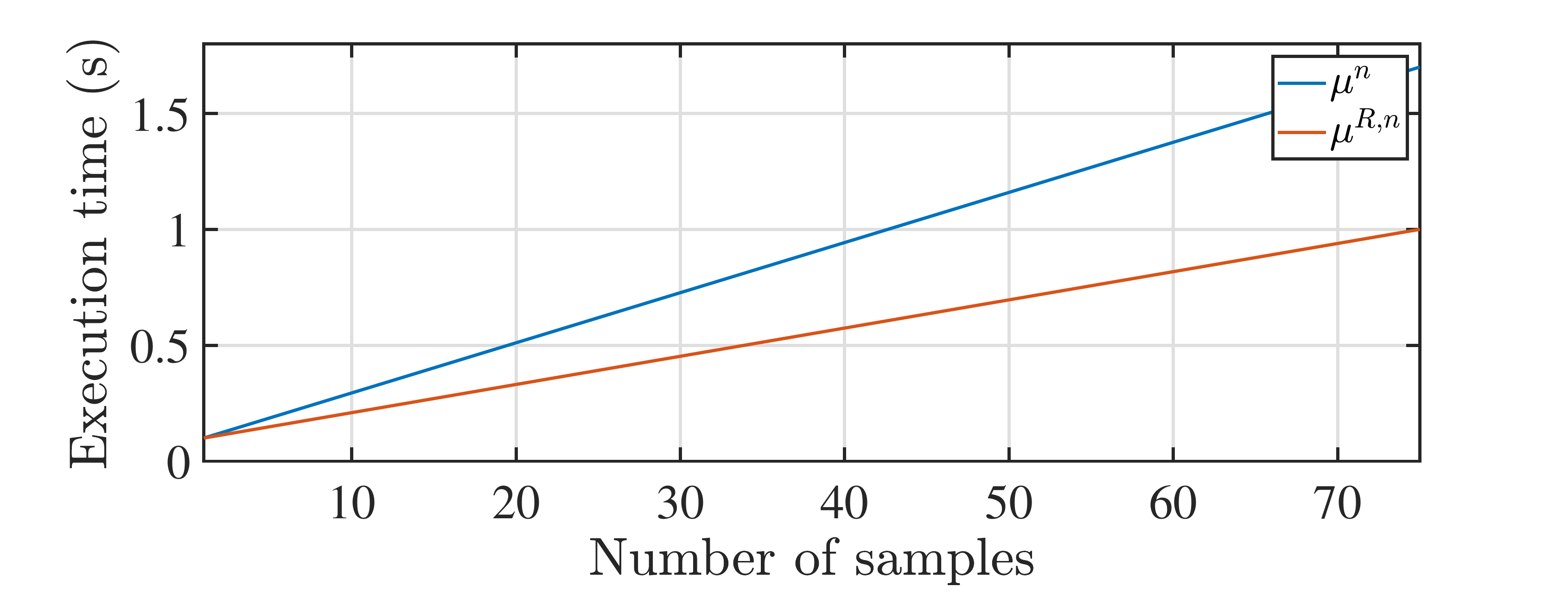}
\caption{Computational time}
\label{fig:time}
\end{subfigure}
\begin{subfigure}[b]{\columnwidth}
\centering
\includegraphics[width=9.5cm]{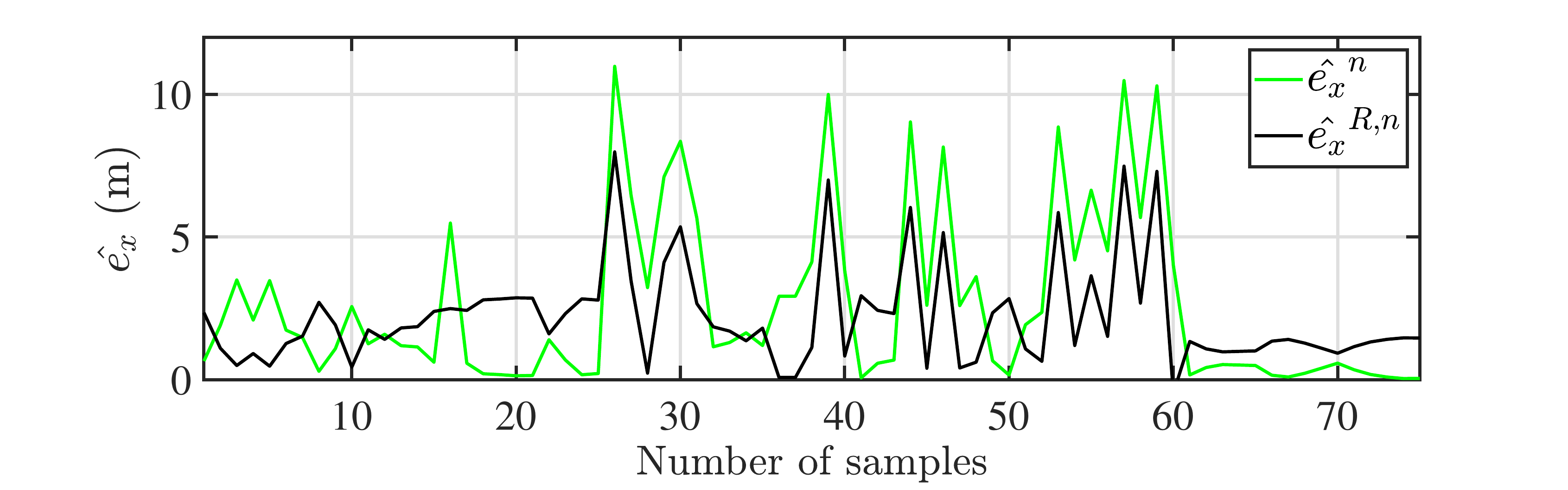}
\caption{$\hat{e}_{x}$}
\label{fig:error}
\end{subfigure}
\caption{(a) Computational time and (b) Estimation error $\hat{e}_{x}$ (in meters) of the $x$-coordinates for the KGCB algs. with and without SP.}
\label{fig:timeanderrorX}
\end{figure}

Furthermore, the performance of the subset policy is evaluated in comparison to the full algorithm execution. As expected, the computational time is significantly reduced (Fig. \ref{fig:time}), with the estimation error $\hat{e}_{x}$ (Fig. \ref{fig:error}) also further reducing when SP is employed, as compared to the results in Fig. \ref{fig:errorXYNL}, since now only the subset of most useful frequency bands are employed. In addition, and as observed previously, the non-linear model that also utilizes the SP achieves better performance results compared to the linear model.

\begin{figure}[t!]
\centering
\includegraphics[width=9.5cm]{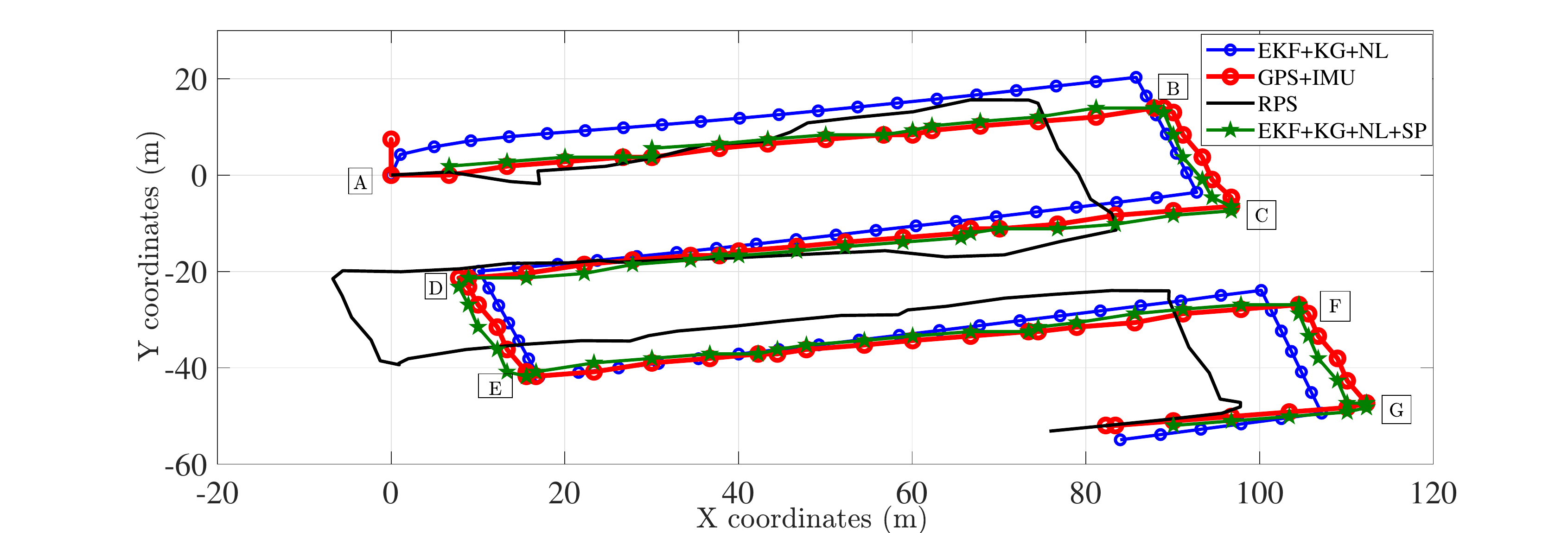}
\caption{Trajectory comparisons while employing GPS+IMU and the different variants of RPS+FSA.}
\label{fig:GPSvsRPStrajectory}
\end{figure}

Finally, Fig.~\ref{fig:GPSvsRPStrajectory} illustrates the vehicle trajectory recorded by GPS+IMU, overlayed by the RPS+FSA trajectories with and without SP. As demonstrated by these trajectories, RPS+FSA manages to accurately follow the GPS+IMU recorded values, while also doing this efficiently and fast (i.e., in real time).

\section{Conclusion}
\label{conclusion}
In this work, a method for accurate and fast positioning is described and evaluated (RPS+FSA), that adaptively chooses a sequence of frequency bands in an online fashion considering RSS values extracted from SOPs and IMU measurements. The problem of choosing which frequency bands to consider is formulated in such a way so as to minimize the positioning error, while also reducing computational time and memory requirements. The algorithm used for calculating the frequency bands that maximize the accuracy of relative localization ($x-y$ coordinates) can be efficiently implemented in scenarios with large RSS datasets. This is demonstrated in an experimental set-up of a practical real-world scenario, that showed that utilizing this method can achieve high positioning accuracy in real time. 

Extensions to this work include fusion of position estimates using information from RGB and TOF cameras, as well as considering distributed techniques with multiple agents that are collaborating with each other for localization purposes.


\section*{Acknowledgment}
This work has been supported by the European Union's Horizon 2020 research and innovation programme under grant agreements No 833611 (CARAMEL) and 739551 (KIOS CoE - TEAMING) and from the Republic of Cyprus through the Deputy Ministry of Research, Innovation and Digital Policy.

\end{document}